\documentclass[aps,prb,showpacs,preprint,floatfix]{revtex4-1}

\usepackage{graphicx}
\usepackage{dcolumn}
\usepackage{amsmath}
\usepackage{rotating}
\usepackage[latin1]{inputenc}

\begin{document}

\title{Magnetic Properties of Pr$_{0.7}$Ca$_{0.3}$MnO$_3$/SrRuO$_3$ Superlattices}
\author{M. Ziese}
\email{ziese@physik.uni-leipzig.de}
\affiliation{Division of Superconductivity and Magnetism, University of Leipzig, D-04103 Leipzig, Germany}
\author{I. Vrejoiu}
\email{vrejoiu@mpi-halle.de}
\affiliation{Max Planck Institute of Microstructure Physics, D-06120 Halle, Germany}
\author{E. Pippel}
\affiliation{Max Planck Institute of Microstructure Physics, D-06120 Halle, Germany}
\author{E. Nikulina}
\affiliation{Max Planck Institute of Microstructure Physics, D-06120 Halle, Germany}
\author{D. Hesse}
\affiliation{Max Planck Institute of Microstructure Physics, D-06120 Halle, Germany}

\date{\today}

\begin{abstract}
High-quality Pr$_{0.7}$Ca$_{0.3}$MnO$_3$/SrRuO$_3$ superlattices were fabricated by pulsed laser deposition and were investigated by
high-resolution transmission electron microscopy and SQUID magnetometry.
Superlattices with orthorhombic and tetragonal SrRuO$_3$ layers were investigated. The superlattices grew coherently; in the growth direction
Pr$_{0.7}$Ca$_{0.3}$MnO$_3$ layers were terminated by MnO$_2$- and SrRuO$_3$ layers by RuO$_2$-planes. All superlattices showed
antiferromagnetic interlayer coupling in low magnetic fields. The coupling strength was significantly higher for orthorhombic than for
tetragonal symmetry of the SrRuO$_3$ layers. The strong interlayer exchange coupling in the superlattice with orthorhombic SrRuO$_3$ layers
led to a magnetization reversal mechanism with a partially inverted hysteresis loop.
\end{abstract}
\maketitle
\clearpage

Epitaxial heterostructures and superlattices (SLs) of perovskite oxides with different physical properties are an exciting playground,
often leading to outstanding physical behavior which cannot be met in the individual compounds. For example, the intriguing magnetic
interlayer coupling between manganites, such as La$_{0.7}$Sr$_{0.3}$MnO$_3$ (LSMO), and SrRuO$_3$ (SRO) in epitaxial bilayers and SLs
were studied both experimentally and theoretically.\cite{ziese2010a} From {\em ab initio} calculations the antiferromagnetic (AF) interlayer
coupling between LSMO and SRO was shown to be mediated by the Mn-O-Ru bond;\cite{lee2008b,ziese2010a}
the AF interlayer coupling depended sensitively on interfacial intermixing and could be controlled using the intricate interplay between structure,
magnetocrystalline anisotropy, magnitude of the layer magnetization and layer thickness.\cite{ziese2010a,ke2004,ke2005,padhan2006}
Additionally, Weigand {\em et al.} reported that the Ru sublattice has
antiferromagnetic coupling to the Mn sublattice in the doped Ruddlesden-Popper phase La$_{1.2}$Sr$_{1.8}$Mn$_{2-x}$Ru$_x$O$_7$.\cite{weigand2002}
Having in mind this tendency of antiferromagnetic coupling between the manganese perovskite R$_{1-x}$A$_x$MnO$_3$ systems
(R = La or rare earth element, and A = Ca, Sr or Ba) and the ruthenates, we investigated the magnetic
interlayer coupling in superlattices in which SRO was combined with Pr$_{0.7}$Ca$_{0.3}$MnO$_3$ (PCMO).
Our PCMO films undergo a single magnetic transition into an insulating ferromagnetic or canted antiferromagnetic state
\cite{jirak1985,yoshizawa1995,tomioka1996} with a Curie temperature of about 110~K, lower than the Curie temperature of the SRO
layers of 160~K.\cite{klein1996a} Both SRO and PCMO have orthorhombic structure in the bulk. The present work
was mainly motivated by our discovery of a structural transition of the SRO layers from orthorhombic to tetragonal symmetry
as a function of PCMO layer thickness,\cite{ziesePCMO} since this opens up the possibility to study exchange coupling
in the same system, but for different structural symmetry.

Superlattices and single films were fabricated by pulsed laser deposition (248~nm, KrF laser) from polycrystalline targets.
Substrate temperature was 650$^\circ$C and oxygen partial pressure 0.14~mbar.
Vicinal SrTiO$_3$$(001)$ substrates with a miscut angle of about $0.1^\circ$, uniform TiO$_2$--termination and an atomically flat
terrace morphology were used. The microstructure of the SLs was investigated by transmission electron microscopy (TEM), atomic
force microscopy and X-ray diffractometry. High-angle annular dark-field scanning transmission electron microscopy (HAADF-STEM), electron energy loss spectroscopy
(EELS) and energy dispersive X-ray (EDX) mappings were done in a TITAN 80-300 FEI microscope with a spherical aberration corrected ($c_s = 0$)
probe forming system.
The magnetic properties of the SLs were investigated by SQUID magnetometry.
The magnetic moments were normalized to the volume of the samples and were expressed in Bohr
magneton per unit cell using an average unit cell size of $(0.39$~nm$)^3$.
Three SLs and two single PCMO films were studied, see table~\ref{table1} for an overview.

Figure~\ref{s1} shows HAADF-STEM micrographs of PCMO/SRO SLs (a) SL1 and (b) SL3. The interfaces between the PCMO and SRO layers were coherent,
no misfit dislocations were found. Closer inspection of the HAADF-STEM micrographs revealed an asymmetry of the interfaces: in the growth direction,
the PCMO layers terminate most probably with MnO$_2$ planes and the SRO layers with RuO$_2$ planes.
As discussed elsewhere\cite{ziesePCMO} the PCMO layers were orthorhombic in all SLs studied here; the
SRO layers, however, were orthorhombic in SL1, whereas having mainly tetragonal symmetry in SL2 and SL3.
Bulk PCMO and SRO have orthorhombic structures at room temperature, however, for epitaxial films, especially coherent and ultrathin ones,
grown on dissimilar substrates, distortions from the orthorhombic bulk structure and formation of particular configurations of crystallographic
domains are expected to occur.\cite{fujimoto2007,gan1999} With respect to the cubic SrTiO$_3$$(001)_C$ substrate the orthorhombic SRO layers
had an epitaxial relation with $[110]_O\parallel[001]_C$, $[001]_O\parallel [100]_C$, $[1\overline{1}0]_O\parallel[010]_C$, and the
tetragonal SRO layers with $[001]_T\parallel[001]_C$, $[1\overline{1}0]_T\parallel[100]_C$, $[110]_T\parallel[010]_C$.
The PCMO layers grew with $[110]_O$ along the substrate normal, but showed two crystallographic domains with the $c$-axis either along $[100]_C$ or $[010]_C$.

Figure~\ref{m1}(a) shows the magnetization vs. temperature curves of both the 35 and 5~nm thick PCMO single
films. The magnetization was measured on field cooling (FCC) in an in-plane field of 0.01~T, respectively, as well as in remanence during warming 
after removal of the magnetic field (REM). In the weak magnetic field of 0.01~T both samples
show a small ferromagnetic-like magnetic moment below about 115~K, although the magnetization onset is very gradual in case of the 5~nm
thick PCMO film. The Curie temperature was determined from the onset of the FCC magnetization measured in 0.01~T and was
found to be about 115~K for the 35~nm and in the range between 110-115~K for the 5~nm thick PCMO film.
The remanent magnetization at low temperature is a sizeable fraction of the FCC magnetization, since
the magnetically hard axis is along the surface normal.

Fig.~\ref{m1}(b) shows the magnetization hysteresis loops
of both films after correcting for the diamagnetic substrate contribution and the paramagnetic contribution
from the Pr ions. Both films show a ferromagnetic signal; saturation is reached only in large fields of the order of 2~T. Compared
to the spin-only moment of Mn of 3.7~$\mu_B$/u.c. both films show reduced saturation moments in agreement with [\onlinecite{yoshizawa1995}].
In summary, the PCMO films studied here show weak ferromagnetism with a Curie temperature between 110 and 115~K, i.e.~lower than the Curie
temperature of the SRO layers of about 143~K.

The low field (0.001~T) magnetization curves of the SLs are shown in Fig.~\ref{m2} as a function of temperature. The
magnetization curves of SL2 and SL3 with tetragonal SRO layers are similar with a dip at about 110~K. Since
this feature is absent in the magnetization of both PCMO single films as well as SRO single films,\cite{zieseprb2010}
we interpret this magnetization dip as a signature of AF interlayer coupling between the SRO and PCMO layers. On cooling
the SLs, first the SRO layers order ferromagnetically below about 143~K; below about 115~K the PCMO layers also start to order ferromagnetically,
but orient their magnetization antiparallel to the SRO magnetization, thus causing a dip in the overall
magnetic moment. In Fig.~\ref{m2}(c) the magnetization of SL1 with orthorhombic SRO layers is shown.
Here the phenomenology is somewhat different, since a peak is observed in the overall magnetic moment indicating
the onset of the AF interlayer coupling. SL1 showed a magnetization maximum both when measured along the magnetically hard $[001]_O$ and easy
$[1\overline{1}0]_O$ in-plane direction, albeit the magnetization values were strongly and the maximum temperatures were slightly different.
This proves that the maximum is not due to a specific magnetization process.

Having observed AF interlayer coupling in the PCMO/SRO superlattices in low magnetic fields, a
qualitative measure of the AF coupling strength was obtained by measuring the temperature-dependent magnetization in higher
applied fields (not shown). Whereas SL2 and SL3 with tetragonal SRO layers do not show
any sign of AF coupling in magnetic fields of 0.1~T and above, the maximum in the magnetization of SL1
persists up to at least 1~T. This indicates that the AF interlayer coupling is stronger in the SL with orthorhombic SRO layers
than in those with tetragonal SRO layers.

The latter conclusion is corroborated by the magnetization hysteresis loops shown in Fig.~\ref{m4}. The hysteresis
loops for in-plane applied fields show a magnetically soft (PCMO) and hard (SRO) component in case of SL2
and SL3; the magnetization reversal occurs in a two-step process with the magnetically soft layer reversing first
and with no obvious indication of a strong AF coupling. In contrast, in SL1 with orthorhombic SRO layers
the in-plane hysteresis loop has an inverted central part, i.e.~the PCMO layer that is reversing first is so strongly
exchange coupled to the SRO layers that the reversal already occurs for positive fields. This inverted hysteresis loop
was also observed for a LSMO/SRO SL\cite{zieseapl} and for the exchange-spring system DyFe$_2$/YFe$_2$.\cite{dumesnil2005}
The out-of-plane magnetization loops show some hysteresis indicating that the magnetic easy axis lies
under some angle with respect to the film plane.

In summary, high-quality PCMO/SRO superlattices and single PCMO thin films were grown on SrTiO$_3$
substrates by pulsed laser deposition. The PCMO single films show ferromagnetic behaviour with a Curie temperature
in the range 110 to 115~K depending on film thickness. Antiferromagnetic interlayer exchange coupling was observed in the SLs
in low magnetic fields. The strength of the AF interlayer exchange coupling depends on the crystallographic symmetry
of the SRO layers and is larger for orthorhombic SRO than for tetragonal SRO layers. This is consistent with the magnetization
reversal mechanism observed in the SLs that leads to inverted hysteresis loops in case of strong AF interlayer exchange coupling.
The structural transition of the SRO layers changes the magnetocrystalline anisotropy and therefore also the spin structure at the interface.
Probably the Ru and Mn spins at the interface are more collinear in case of orthorhombic SRO symmetry leading to a stronger AF exchange coupling.

\acknowledgments This work was supported by the German Science Foundation (DFG) within the Collaborative Research Center SFB 762
``Functionality of Oxide Interfaces''. We thank Dr.~J.~Henk for a careful reading of the manuscript.

\clearpage

%\bibliography{PCMO_M}
%

\clearpage

\begin{table}
\caption{Layer thicknesses and Curie temperatures of the SLs and single PCMO films. Curie temperatures of the PCMO layers
were determined from magnetization, those of the SRO layers from resistivity measurements.}
\begin{tabular}{lccc}
\hline
Sample & $[$PCMO / SRO$]_{15}$ & $T_C$ (K) & $T_C$ (K)\\
       &                       &   (SRO)   & (PCMO)\\
\hline
SL1 &  [1.5~nm / 4.4~nm] & 143 & $\simeq 110$\\
SL2 &  [3.0~nm / 4.0~nm] & 143 & $\simeq 110$\\
SL3 &  [3.8~nm / 4.0~nm] & 142 & $\simeq 115$\\
\hline
PCMO1 &  5~nm & -- & $\simeq 110-115$\\
PCMO2 &  35~nm & -- & $\simeq 115$\\
\hline
\end{tabular}
\label{table1}
\end{table}

\clearpage
\begin{figure}[t]
\caption{HAADF-STEM images of samples (a) SL1 (1.5~nm/4.4~nm) and (b) SL3 (3.8~nm/4.0~nm).}
\label{s1}
\end{figure}
\begin{figure}[t]
\caption{(a) Magnetization vs. temperature curves of the 35~nm (solid symbols)
and 5~nm (open symbols) thick single PCMO films for a field of 0.01~T applied in-plane.
Field-cooled (FCC) and remanence (REM) data are shown.
(b) In-plane magnetization hysteresis loops of the 35~nm and 5~nm single PCMO films at 10~K.}
\label{m1}
\end{figure}
\begin{figure}[t]
\caption{Low-field (0.001~T) magnetization vs.~temperature curves of (a) SL2 and (b) SL3 with tetragonal
as well as (c) SL1 with orthorhombic SRO layers. In (a) and (b) the field was applied in-plane along $[1\overline{1}0]_T$,
in (c) in-plane along $[001]_O$ and $[1\overline{1}0]_O$. For comparison in (b)
the in-plane magnetization of single film PCMO2 is shown. The arrows indicate the onset of AF coupling.
Crystallographic directions are indicated schematically.}
\label{m2}
\end{figure}
\begin{figure}[t]
\caption{Magnetization hysteresis loops of SLs (a) SL2, (b) SL3 and (c) SL1 at 10~K
for both in-plane (solid symbols) and out-of-plane (open symbols) magnetic fields. The arrows in (c)
indicate the sweep direction of the in-plane field; note that the central loop of the in-plane
magnetization curve is inverted. The field directions are indicated in the figure.}
\label{m4}
\end{figure}

\clearpage
\setcounter{figure}{0}
\begin{figure}[t]
\centerline{\includegraphics[width=0.7\textwidth]{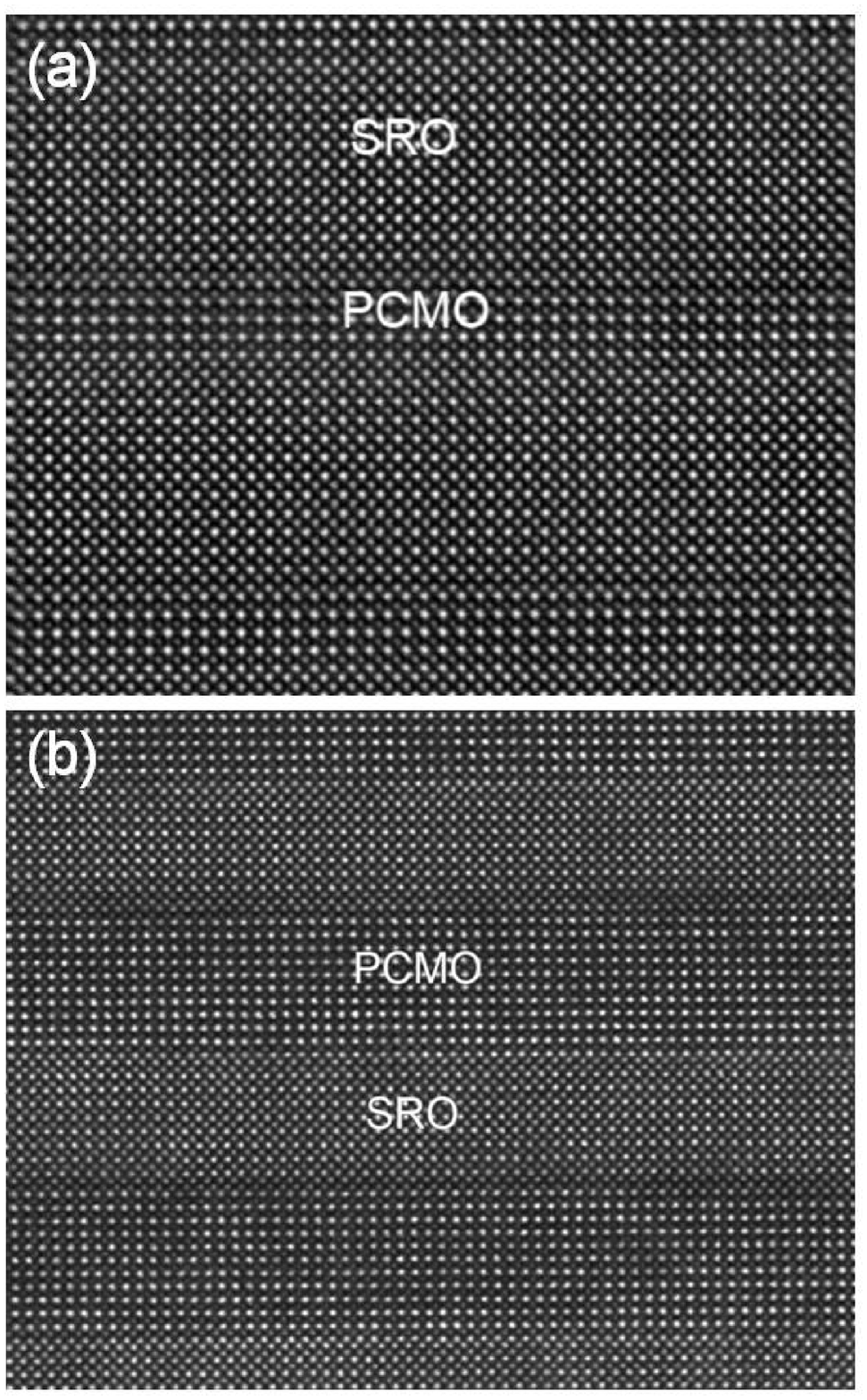}}
\caption{}
\end{figure}
\clearpage
\begin{figure}[t]
\centerline{\includegraphics[width=0.7\textwidth]{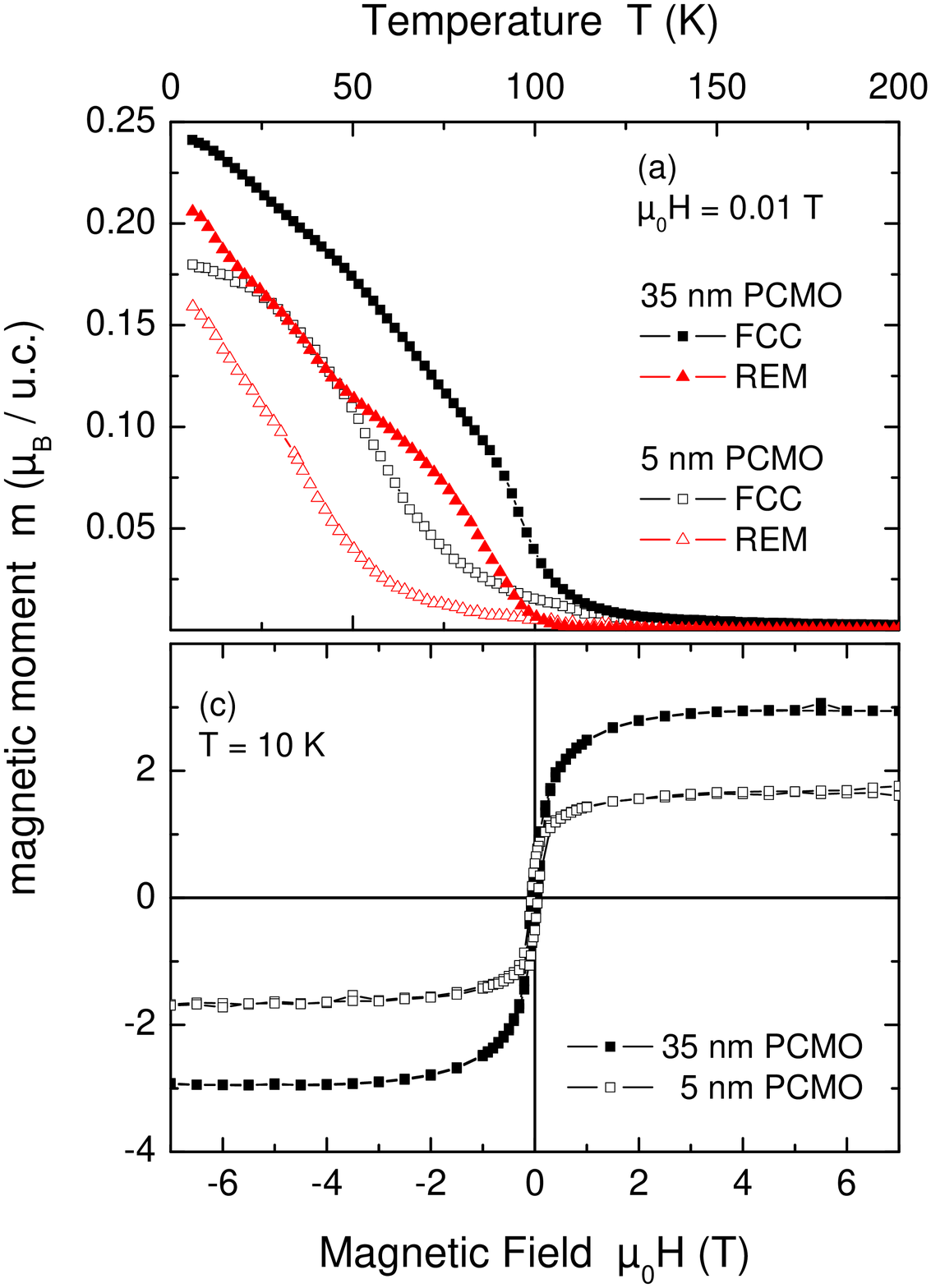}}
\caption{}
\end{figure}
\clearpage
\begin{figure}[t]
\centerline{\includegraphics[width=0.7\textwidth]{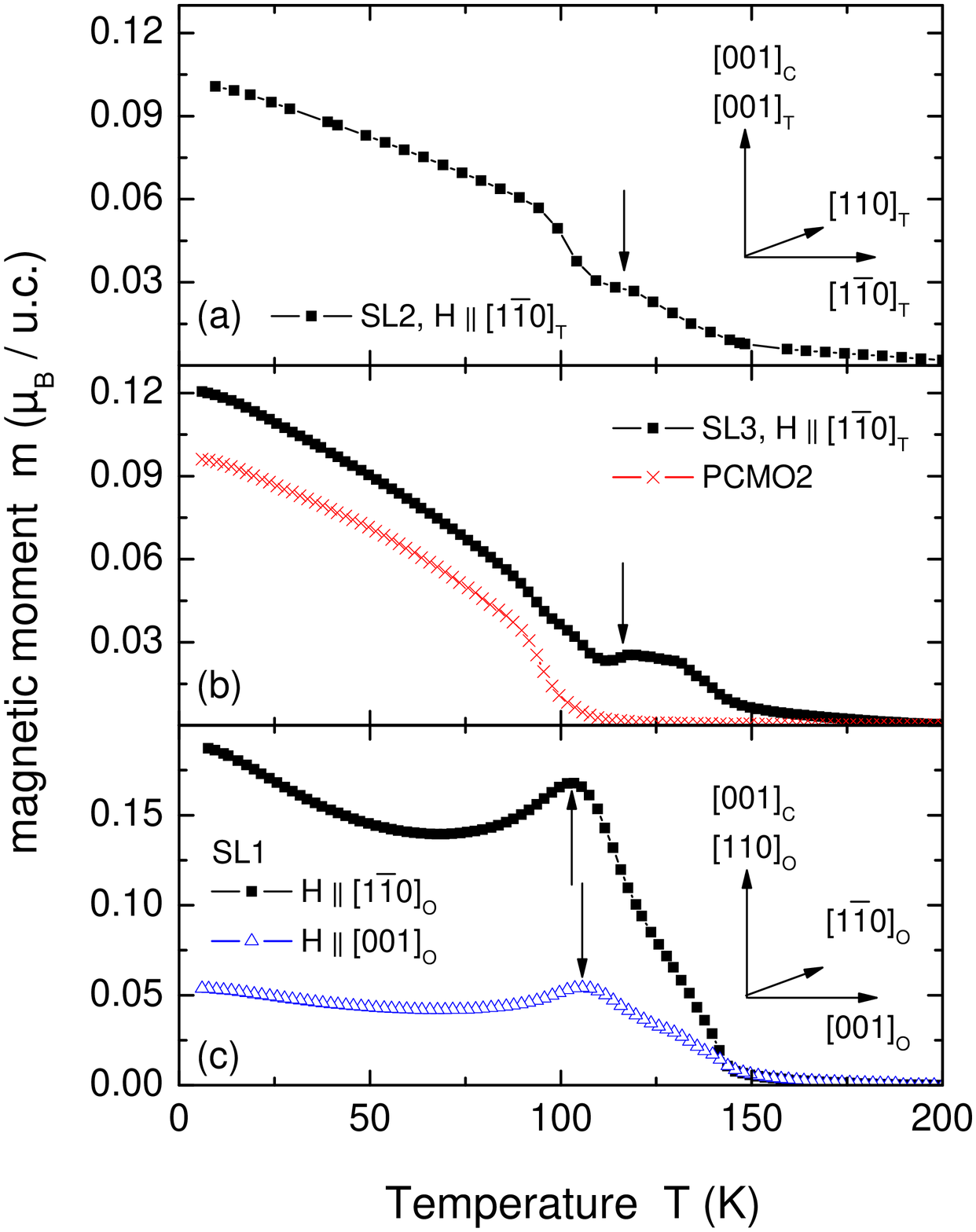}}
\caption{}
\end{figure}
\clearpage
\begin{figure}[t]
\centerline{\includegraphics[width=0.7\textwidth]{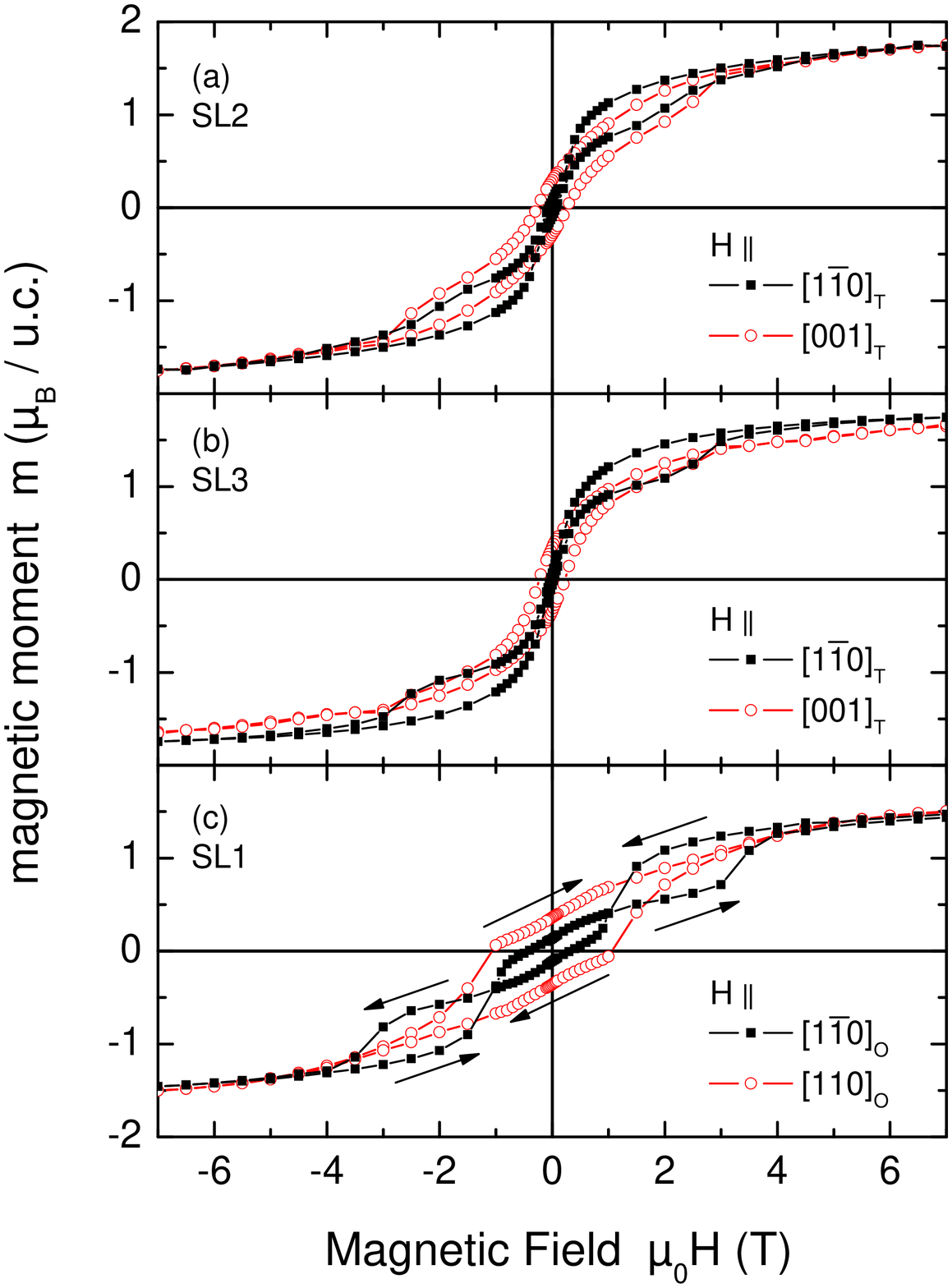}}
\caption{}
\end{figure}

\end{document}